\documentclass[runningheads]{llncs}
\usepackage{physics}
\usepackage{mathptmx}
\usepackage{feynmf}
\usepackage{tikz-feynman}
\usepackage{amsmath}
\usepackage{amssymb}
\usepackage{mathtools}
\usepackage{calrsfs}
\DeclareMathAlphabet{\pazocal}{OMS}{zplm}{m}{n}
\usepackage{graphicx}
\usepackage{graphicx}
\usepackage{caption}
\usepackage{color}
\usepackage{subcaption}
\usepackage{amsmath}
\usepackage{listings}
\usepackage{dirtytalk}
\usepackage{url}
\usepackage{threeparttable}
\usepackage{adjustbox}
\usepackage{wrapfig}
\begin{document}
\title{Discussion on Inert Doublet Model and loop correction of $HH\xrightarrow{}hh$\thanks{University Of Delhi}}
%
%
\author{Sumit Satapathy\inst{1}\orcidID{0000-0003-1146-1251}}
\authorrunning{Satapathy.S}
%
\institute{Department of Physics and Astrophysics, University of Delhi
, New Delhi. 110021 New Delhi, India
\\
}
\maketitle              
\begin{abstract}
In this We take the Lagrangian of Inert Doublet Model and with unitary gauge calculate all the
possible interactions after calculating the interactions we draw the corresponding Feynman diagrams and
there after calculate the Scattering cross section of the processes that contribute in the decay or scattering of
DM candidate.  In the concluding
chapter in which we calculate the one loop correction to
the process $HH\xrightarrow{}hh$ and see all the possible loop contribution at this scale.
\end{abstract}
\section{Inert Doublet Model}
The very first idea of considering the two Higgs model was put forward by Deshpande and Ma\cite{1}. The particle candidate is a weakly interacting massive scalar as discussed in the below chapters. The frame work is that of a two Higgs doublet (2HDM),  {$H_1$}  and  $H_2$, version of the Standard model
with a discrete symmetry of  $Z_2$  such that \cite{2}$$H_1\xrightarrow{}H_1  ,  H_2\xrightarrow{}-H_2 $$
                  
This discrete symmetry can be understood from the $Z_n$ multiplicative symmetry group of roots of unity which  has (-1,+1) only and we define Lagrangian to be either positive or negative under this symmetry.\\
Such that if a process starts from even particles the result should give me $Z_2$ even particles (+1*+1=+1) so as this $Z_2$ symmetry explain very easily why something very heavy might also be stable. Therefore if we start out with an odd BSM particle there is no way it can decay in SM even particle.
So, $Z_2$ are the most convenient way to add DM candidates such that the decays has an extra discrete symmetry to full fill.\\
 The second inert doublet contains two charged and two neutral scalars and as they are odd under the imposed symmetry, its lightest neutral component provides a natural candidate for dark matter (DM).\cite{3}\\ 
 \subsection{Lagrangian of IDM}
After imposing the symmetry under discrete transformation, the full scalar potential can be written as,
$$V=m_1^2{|H_1|^2}+m_2^2{|H_2|^2}+\lambda_1{|H_1|^4}+\lambda_2{|H_2|^4}+\lambda_3{|H_1|^2|H_2|^2}+\lambda_4{(H_1^+H_2)(H_2^+H_1)}+$$ $$\frac{\lambda_5}{2}{((H_1^+H_2)^2+ h.c)}$$

This is the most general $SU(2)\cross U(1)$ gauge invariant, renormalizable higgs potential for two doublet\cite{4}. The $H_1$ belongs to the higgs field of standard model which undergoes symmetry break down, and our $H_2$ doesn't participate in electroweak symmetry breaking or we can say that the vev is non-zero in case of standard model field. The identification of DM candidate can be made, as the second higgs doublet has charged and neutral components we will see through the Feynman diagrams that the lightest neutral component provides a natural candidate for dark matter. Moreover, due to the discrete symmetry the lightest neutral scalar cannot decay therefore provide a candidate for dark matter.\cite{5}\cite{6}\cite{7}\\
In standard model the  $SU(2)_L\cross U(1)_Y$  symmetry is broken by vev of complex doublet $H$ with hypercharge $\frac{1}{2}$  called higgs multiplet. In addition to this symmetry we have also added that the second higgs has a global $U(1)$ symmetry with $Z_2$ parity.The fact that the extra doublet has a nonzero 
charge implies that its vacuum expectation value is vanishing\
After we use the unitary gauge, In this gauge, the scalar fields responsible for the Higgs mechanism are transformed into a basis in which their Goldstone boson components are set to zero.
 
\centerline{$H_1$=$
\begin{pmatrix}
    0\\
    \frac{1}{\sqrt{2}}(v+h)\\
\end{pmatrix}
$          \space\space\space                    $H_2$=$\begin{pmatrix}
    H^+\\
    \frac{1}{\sqrt{2}}(H+iA)\\
\end{pmatrix}
$}
\textbf{The interaction vertices are}:
$$H_1^+H_1\\
=\begin{pmatrix}
    0 & \frac{1}{\sqrt{2}}(v+h)
\end{pmatrix}
\begin{pmatrix}
    0\\
    \frac{1}{\sqrt{2}}(v+h)
\end{pmatrix}
$$
$$=\frac{1}{2}(v^2+h^2+2vh)$$
$$H_2^+H_2
=\begin{pmatrix}
    H^- & \frac{1}{\sqrt{2}}(H-iA)
\end{pmatrix}
\begin{pmatrix}
    H^+\\
    \frac{1}{\sqrt{2}}(H+iA)
\end{pmatrix}
$$
$$=H^-H^++\frac{1}{2}(H^2+A^2)$$\\
$$(H_1^+H_2)(H_2^+H_1)
=\begin{pmatrix}
    0 & \frac{1}{\sqrt{2}}(v+h)
\end{pmatrix}
\begin{pmatrix}
    H^+\\
    \frac{1}{\sqrt{2}}(H+iA)
\end{pmatrix}
\begin{pmatrix}
    H^- & \frac{1}{\sqrt{2}}(H-iA)
\end{pmatrix}
\begin{pmatrix}
    0\\
    \frac{1}{\sqrt{2}}(v+h)
\end{pmatrix}$$
$$=\frac{1}{4}(v^2H^2+v^2A^2+h^2H^2+h^2A^2+2vhH^2+2vhA^2)$$\\
$$(H_1^+H_1)(H_2^+H_2)
=\begin{pmatrix}
    0 & \frac{1}{\sqrt{2}}(v+h)
\end{pmatrix}
\begin{pmatrix}
    0\\
    \frac{1}{\sqrt{2}}(v+h)
\end{pmatrix}
\begin{pmatrix}
    H^- & \frac{1}{\sqrt{2}}(H-iA)
\end{pmatrix}
\begin{pmatrix}
    H^+\\
    \frac{1}{\sqrt{2}}(H+iA)
\end{pmatrix}$$
$$=\frac{1}{2}(v+h)^2\left(H^-H^++\frac{1}{2}(H^2+A^2)\right)$$\\
$$(H_1^+H_2)
=\begin{pmatrix}
    0 & \frac{1}{\sqrt{2}}(v+h)
\end{pmatrix}
\begin{pmatrix}
    H^+\\
    \frac{1}{\sqrt{2}}(H+iA)
\end{pmatrix}$$
$$=\frac{1}{2}(vH+iAv+hH+iAh)$$
These are the possible interaction terms possible that are seen from the potential term. Now we write the entire potential.\\
$$V=\left(\frac{m_1^2v^2}{2}+\frac{m_1^2h^2}{2}+m_1^2vh\right)+\left(m_2^2H^-H^++\frac{m_2^2H^2}{2}+\frac{m_2^2A^2}{2}\right)+$$$$\frac{\lambda{}_1}{2}\left(v^4+h^4+4v^2h^2+2v^2h^2+2v^3h+2vh^3\right)+\lambda{}_2(H^-H^+H^-H^++\frac{H^4}{4}+\frac{A^4}{4}+H^-H^+HH+$$
$$+H^-H^+AA+\frac{HHAA}{2})+\frac{\lambda{}_4}{4}\left(v^2HH+v^2AA+hhHH+hhAA+2vhHH+2vhAA\right)+$$$$\frac{\lambda{}_3}{2}\left(v^2H^-H^++\frac{v^2}{2}(HH+AA)+hhH^-H^++2vhH^-H^++\frac{h^2}{2}(HH+AA)+vh(HH+AA)\right)+$$$$\frac{\lambda{}_5}{8}\left(v^2HH-AAv^2+hhHH-AAhh+2vHhH-2vAAh\right)$$
\\
 the possible scalar vertices are\cite{3},\\\\
\begin{table}[h]
    \centering
    \begin{tabular}{|c|c|}
    \hline
   Quartic scalar vertices & Triple scalar vertices\\\hline
   $hhhh$  & $hhh$ \\\hline
   $hhH^-H^+$  &  $hH^-H^+$\\\hline
   $H^-H^+H^-H^+$ & $hHH$\\\hline
   $HHHH$ & $hAA$\\\hline
   $AAAA$ & \\\hline
   $H^-H^+HH$ & \\\hline
   $H^-H^+AA$ & \\\hline
   $HHAA$ & \\\hline
   $hhHH$ & \\\hline
   $hhAA$ & \\\hline
   
    \end{tabular}
    \caption{Scalar vertices}
    \label{tab:my_label}
\end{table}
 You might worry and think, what about the coupling constants, the new coupling constants will we obtained when we take common the above vertices in the potential and we get some new coupling constants in form of, sum or differences of the old \cite{2}ones($\lambda_1{},\lambda_2{},\lambda_3{},\lambda_4{},\lambda_5$).\\
For the inclusion of gauge scalar vertices we have the term of,
$$D^\mu\phi^+_1D_\mu\phi_1 , \space \space D^\mu\phi^+_2D_\mu\phi_2$$
The covariant derivative term,$$D_\mu=\partial_\mu+ige.W_\mu-i\frac{g'}{2}Y_\mu$$

$$\Phi_2\xrightarrow{}\begin{pmatrix}
    H^+ \\
    \frac{1}{\sqrt{2}}(H+iA)
\end{pmatrix}$$\\
$$\frac{\sigma}{2}.W^a_\mu=\frac{1}{2}\begin{pmatrix}
    W^3 & \sqrt{2}W^+\\
    \sqrt{2}W^- & -W^3\\
\end{pmatrix}
    $$
     Where,
    $$W^+=\frac{1}{\sqrt{2}}(W^1-iW^2)$$
    $$W^-=\frac{1}{\sqrt{2}}(W^1+iW^2)$$\\
 Now the term $D^\mu\Phi^+_2 D_\mu\Phi_2$, Becomes,
 $$D^\mu\Phi^+_2 D_\mu\Phi_2=(\partial^\mu+ig\frac{\sigma}{2}.W^a-i\frac{g'}{2}Y)(\Phi^+_2)(\partial_\mu+ig\frac{\sigma}{2}.W^a-i\frac{g'}{2}Y)(\Phi_2)$$
$$=\partial^\mu\Phi^+_2\partial_\mu\Phi_2+ig\partial^\mu\Phi^+_2\frac{\sigma}{2}.W\Phi_2+\frac{ig'}{2}\partial^\mu\Phi^+_2Y\Phi_2$$ 

 $$+\left(ig\frac{\sigma}{2}.W\Phi^+_2\partial_\mu\Phi_2\right)+\left(i\frac{g}{2}\sigma.W\Phi^+_2ig\frac{\sigma}{2}.W\Phi_2\right)+\left(i\frac{g}{2}\sigma.W\Phi^+_2i\frac{g'}{2}Y\Phi_2\right)$$ $$+\left(i\frac{g'}{2}Y\Phi^+_2\partial_mu\Phi_2\right)+\left(i\frac{g'}{2}Y\Phi^+_2i\frac{g}{2}\frac{\sigma}{2}.W\Phi_2\right)+\left(\frac{ig'}{2}Y\Phi^+_2i\frac{g'}{2}Y\Phi_2\right)$$\\

 $$=\begin{pmatrix}
   
    W^3 & \sqrt{2}W^+\\
    \sqrt{2}W^- & -W^3\\
 \end{pmatrix}\begin{pmatrix}
     H^+ \\
    \frac{1}{\sqrt{2}}(H+iA)
 \end{pmatrix}
 $$\\
 $$\xrightarrow{}\begin{pmatrix}
     W^3H^+ + W^+H + W^+iA\\
     \sqrt{2}W^-H^+ - \frac{W^3}{\sqrt{2}H}-\frac{W^3iA}{\sqrt{2}}\\
 \end{pmatrix}$$
 $$\xrightarrow{}
 \begin{pmatrix}
     \partial^\mu H^- & \frac{1}{\sqrt{2}}(\partial^\mu H-i\partial^\mu A)\\
 \end{pmatrix}\begin{pmatrix}
     W^3H^+ + W^+H + W^+iA\\
     \sqrt{2}W^-H^+ - \frac{W^3}{\sqrt{2}H}-\frac{W^3iA}{\sqrt{2}}\\
 \end{pmatrix}
 $$\\
 $$=(\partial^\mu H^-)W^3H^++(\partial^\mu H^-)W^3H+(\partial^\mu H)^-W^+iA+(\partial^\mu H)W^-H^+ +iA(\partial^\mu A)W^-H^+ $$$$-\frac{1}{2}(\partial^\mu H)W^3H+\frac{i}{2}(\partial^\mu A)W^3H+\frac{i}{2}(\partial^\mu H)W^3A+\frac{1}{2}(\partial^\mu A)AW^3$$\\
 Now similarly as above we can calculate the vertices possible for the entire $D^\mu\phi^+_2D_\mu\phi_2$ term and find out the possible 4-point and triple scalar vertices possible. 
 Corresponding gauge scalar vertices are listed below\cite{3},\\
\begin{table}[h]
    \centering
    \begin{tabular}{|c|c|}
    \hline
        Vertex &  Coupling\\\hline
         $H^-H^+\gamma$ & i e\\\hline
         $H^-H^+Z$ & i$\frac{g}{2}\frac{cos(2\theta_W)}{cos\theta_W}$\\\hline
         $HH^\pm W^\mp$ & $\mp i\frac{g}{2}$\\\hline
         $AH^\mp W^\pm$ & $-\frac{g}{2}$\\\hline
         $HAZ$ & $-\frac{g}{2cos\theta_W}$\\\hline
         \end{tabular}
    \caption{Gauge scalar vertices}
    \label{Gauge scalar verticesl}
\end{table}
$\theta_W$ is known as the Weinberg angle\cite{8},
 $$cos\theta_W=\frac{g}{\sqrt{g^2+g'^2}}$$
$$sin\theta_W=\frac{g'}{\sqrt{g^2+g'^2}}$$
\subsection{Feynman Diagrams}
$$\begin{tikzpicture}
  \begin{feynman}
    \vertex(a);
    \vertex[ above left=of a](b){$A$};
    \vertex[ below left=of a](c){$A$};
    \vertex[ above right=of a](d){$A$};
    \vertex[ below right=of a](e){$A$};
    \diagram{
    (a) -- (b);
    (a) -- (e);
    (a) -- (c);
    (a) -- (d);
             }
    ;
    \end{feynman}
\end{tikzpicture}
\begin{tikzpicture}
  \begin{feynman}
    \vertex(a);
    \vertex[ above left=of a](b){$H$};
    \vertex[ below left=of a](c){$H$};
    \vertex[ above right=of a](d){$H$};
    \vertex[ below right=of a](e){$H$};
    \diagram{
    (a) -- (b);
    (a) -- (e);
    (a) -- (c);
    (a) -- (d);
    }
    ;
    \end{feynman}
\end{tikzpicture}$$\\
$$\begin{tikzpicture}
  \begin{feynman}
    \vertex(a);
    \vertex[ above left=of a](b){$h$};
    \vertex[ below left=of a](c){$h$};
    \vertex[ above right=of a](d){$h$};
    \vertex[ below right=of a](e){$h$};
    \diagram{
    (a) -- (b);
    (a) -- (e);
    (a) -- (c);
    (a) -- (d);
    }
    ;
    \end{feynman}
\end{tikzpicture}
\begin{tikzpicture}
  \begin{feynman}
    \vertex(a);
    \vertex[ above left=of a](b){$A$};
    \vertex[ below left=of a](c){$A$};
    \vertex[ above right=of a](d){$A$};
    \vertex[ below right=of a](e){$A$};
    \diagram{
    (a) -- (b);
    (a) -- (e);
    (a) -- (c);
    (a) -- (d);
    }
    ;
    \end{feynman}
\end{tikzpicture}$$\\
$$\begin{tikzpicture}
  \begin{feynman}
    \vertex(a);
    \vertex[ above left=of a](b){$H^+$};
    \vertex[ below left=of a](c){$H^-$};
    \vertex[ above right=of a](d){$H^+$};
    \vertex[ below right=of a](e){$H^-$};
    \diagram{
    (a) -- (b);
    (a) -- (e);
    (a) -- (c);
    (a) -- (d);
    }
    ;
    \end{feynman}
\end{tikzpicture}
\begin{tikzpicture}
  \begin{feynman}
    \vertex(a);
    \vertex[ above left=of a](b){$H$};
    \vertex[ below left=of a](c){$H$};
    \vertex[ above right=of a](d){$A$};
    \vertex[ below right=of a](e){$A$};
    \diagram{
    (a) -- (b);
    (a) -- (e);
    (a) -- (c);
    (a) -- (d);
    };
    \end{feynman}
\end{tikzpicture}$$\\
$$\begin{tikzpicture}
  \begin{feynman}
    \vertex(a);
    \vertex[ above left=of a](b){$H^+$};
    \vertex[ below left=of a](c){$H^-$};
    \vertex[ above right=of a](d){$A$};
    \vertex[ below right=of a](e){$A$};
    \diagram{
    (a) -- (b);
    (a) -- (e);
    (a) -- (c);
    (a) -- (d);
    };
    \end{feynman}
\end{tikzpicture}
\begin{tikzpicture}
  \begin{feynman}
    \vertex(a);
    \vertex[ above left=of a](b){$H^+$};
    \vertex[ below left=of a](c){$H^-$};
    \vertex[ above right=of a](d){$H$};
    \vertex[ below right=of a](e){$H$};
    \diagram{
    (a) -- (b);
    (a) -- (e);
    (a) -- (c);
    (a) -- (d);
    };
    \end{feynman}
\end{tikzpicture}$$\\
$$\begin{tikzpicture}
  \begin{feynman}
    \vertex(a);
    \vertex[ above left=of a](b){$h$};
    \vertex[ below left=of a](c){$h$};
    \vertex[ above right=of a](d){$H^+$};
    \vertex[ below right=of a](e){$H^-$};
    \diagram{
    (a) -- (b);
    (a) -- (e);
    (a) -- (c);
    (a) -- (d);
    };
    \end{feynman}
\end{tikzpicture}
\begin{tikzpicture}
  \begin{feynman}
    \vertex(a);
    \vertex[ above left=of a](b){$h$};
    \vertex[ below left=of a](c){$h$};
    \vertex[ above right=of a](d){$H$};
    \vertex[ below right=of a](e){$H$};
    \diagram{
    (a) -- (b);
    (a) -- (e);
    (a) -- (c);
    (a) -- (d);
    };
    \end{feynman}
\end{tikzpicture}$$\\
$$\begin{tikzpicture}
  \begin{feynman}
    \vertex(a);
    \vertex[ above left=of a](b){$h$};
    \vertex[ below left=of a](c){$h$};
    \vertex[ above right=of a](d){$A$};
    \vertex[ below right=of a](e){$A$};
    \diagram{
    (a) -- (b);
    (a) -- (e);
    (a) -- (c);
    (a) -- (d);
    };
    \end{feynman}
\end{tikzpicture}$$\\
$$\begin{tikzpicture}
  \begin{feynman}
    \vertex(a){$h$};
    \vertex[  right =of a](b);
    \vertex[  above right =of b](c){$A$};
    \vertex[ below right=of b](d){$A$};
    \diagram{
    (a) -- (b);
    (b) -- (c);
    (b) -- (d);
    };
    \end{feynman}
\end{tikzpicture}\\
\begin{tikzpicture}
  \begin{feynman}
    \vertex(a){$h$};
    \vertex[  right =of a](b);
    \vertex[  above right =of b](c){$H$};
    \vertex[ below right=of b](d){$H$};
    \diagram{
    (a) -- (b);
    (b) -- (c);
    (b) -- (d);
    };
    \end{feynman}
\end{tikzpicture}\\
\begin{tikzpicture}
  \begin{feynman}
    \vertex(a){$h$};
    \vertex[  right =of a](b);
    \vertex[  above right =of b](c){$h$};
    \vertex[ below right=of b](d){$h$};
    \diagram{
    (a) -- (b);
    (b) -- (c);
    (b) -- (d);
    };
    \end{feynman}
\end{tikzpicture}\\
\begin{tikzpicture}
  \begin{feynman}
    \vertex(a){$h$};
    \vertex[  right =of a](b);
    \vertex[  above right =of b](c){$H^+$};
    \vertex[ below right=of b](d){$H^-$};
    \diagram{
    (a) -- (b);
    (b) -- (c);
    (b) -- (d);
    };
    \end{feynman}
\end{tikzpicture}$$
\\
These are the kinematically possible diagrams for the vertices given above.
The masses are as follow\cite{9},
$$m^2_h=2\lambda_1{} v^2$$
$${m^2_H}^{\pm}=\lambda_3v^2-m_2^2$$
$$m^2_A={m^2_H}^{\pm}+(\lambda_4-\lambda_5)v^2$$
$$m^2_H={m^2_H}^{\pm}+\lambda_4v^2+\frac{1}{2}\lambda_5v^2$$
The interesting part where we define our DM character is clear now, as the masses are of the order\cite{3},
$$m_H<m_A,m_H^{\pm}$$ And from the D symmetry we know the lightest neutral candidate cannot decay further and thus provides a suitable candidate\cite{2}\cite{3}\cite{6}.
And its decay diagram are valuable for us for the calculation of relic density and cross sections.
\subsection{Scattering cross section for tree-level process}
Before calculating cross section we should group all the coupling constant regarding all the possible interaction.
And the possible interaction that contributes to the decay or scattering are the 4-point vertices with higgs, vector bosons and as well as diagrams which include the
annihilation of a pair of inert doublet particles via the gauge boson or the Higgs channel into
SM particles\cite{4}, These are momentum dependent due to the derivative term and due to that we know we have a factor of $-ik$ and $+ik$ depending on the incoming or outgoing particles\cite{10}.These momentum dependent terms are small compared to the 4-point vertex interaction. The possible diagrams are as follow,
$$
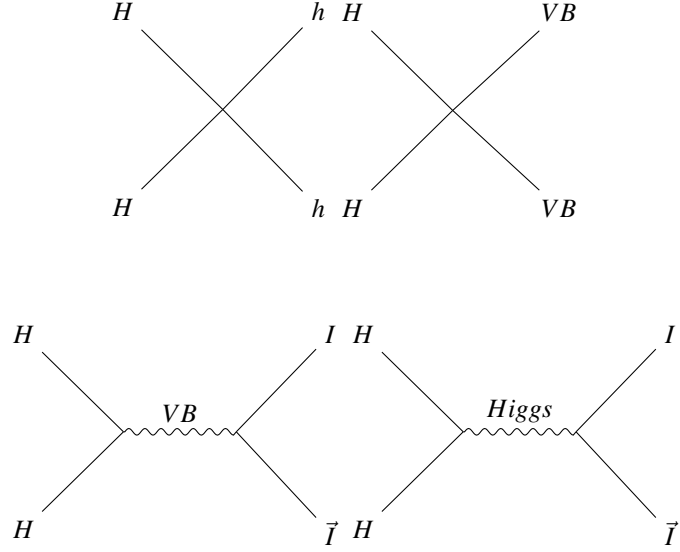
\begin{tikzpicture}
  \begin{feynman}
    \vertex(a);
    \vertex[ above left=of a](b){$H$};
    \vertex[ below left=of a](c){$H$};
    \vertex[ above right=of a](d){$h$};
    \vertex[ below right=of a](e){$h$};
    \diagram{
    (a) -- (b);
    (a) -- (e);
    (a) -- (c);
    (a) -- (d);
    };
    \end{feynman}
\end{tikzpicture}
\begin{tikzpicture}
  \begin{feynman}
    \vertex(a);
    \vertex[ above left=of a](b){$H$};
    \vertex[ below left=of a](c){$H$};
    \vertex[ above right=of a](d){$VB$};
    \vertex[ below right=of a](e){$VB$};
    \diagram{
    (a) -- (b);
    (a) -- (e);
    (a) -- (c);
    (a) -- (d);
    };
    \end{feynman}
\end{tikzpicture}$$\\
$$\begin{tikzpicture}
    \begin{feynman}
        \vertex(a);
        \vertex[above left=of a](c){$H$};
        \vertex[below left=of a ](d){$H$};
        \vertex[right=of a](b);
        \vertex[above right=of b](e){$I$};
        \vertex[below right=of b ](f){$\Vec{I}$};
   
    \diagram{
          (c) -- (a) -- (d);
          (a) -- [boson,edge label=$VB$](b);
          (e) -- (b) -- (f);
    };
    \end{feynman}
\end{tikzpicture}
\begin{tikzpicture}
    \begin{feynman}
        \vertex(a);
        \vertex[above left=of a](c){$H$};
        \vertex[below left=of a ](d){$H$};
        \vertex[right=of a](b);
        \vertex[above right=of b](e){$I$};
        \vertex[below right=of b ](f){$\Vec{I}$};
   
    \diagram{
          (c) -- (a) -- (d);
          (a) -- [boson,edge label=$Higgs$](b);
          (e) -- (b) -- (f);
    };
    \end{feynman}
\end{tikzpicture}$$

Now corresponding to the above the diagrams we have

$$\sigma|_{cm}v_{rel}=\frac{1}{32\pi m_2^2(1+v^2_{rel}/4)}\sum_{processes}|M|^2$$
 $$\sigma|_{cm}v_{rel}=\frac{1}{32\pi m_2^2}\left(1-\frac{v_{rel}^2}{4}\right)\sum_{processes}|M|^2$$
 The amplitude for the neutral scalar component are\cite{4},
$$\sum_{Processes}|M|^2=(\lambda_3{}+\lambda{}_4+\lambda_5)^2+g^4+(g^2+g^{'2})^2+\frac{1}{8}\frac{g^4g^{'2}}{g^2+g^{'2}}+\frac{1}{2}g^2(g^2+g^{'2})$$

 \section{One-Loop correction of $HH\xrightarrow{}hh$}
 We know that at earlier time or when the temperature was much higher, the decay of DM candidate was in equilibrium such that the forward reaction rate and backward reaction rate were same. Now we use the idea of one loop correction, as we know at earlier epoch the dark matter candidate were in equilibrium with the photon bath the corresponding momentum or momentum cutoff should also be high, such that the theory for the decay at the earlier epoch will give finite amplitude only if we re-normalize our theory at some higher momentum.The idea is if we consider that before the dark matter freeze out it has some relativistic properties or "Hot", such that we have to look at higher value momenta in this process we have divergence(UV) so we introduce the method of re-normalization which gives me a new Lagrangian of same form and of same divergence analysis but of redefined fields\cite{11}\cite{12}\cite{13}\cite{14}\cite{15}.  Will see in the coming sections more about this correction to the Lagrangian.
 $$
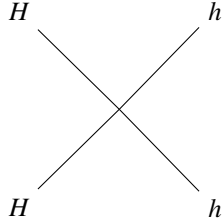
\begin{tikzpicture}
    \begin{feynman}
         \vertex(a);
    \vertex[ above left=of a](b){$H$};
    \vertex[ below left=of a](c){$H$};
    \vertex[ above right=of a](d){$h$};
    \vertex[ below right=of a](e){$h$};
    \diagram{
    (a) -- (b);
    (a) -- (e);
    (a) -- (c);
    (a) -- (d);
    };
    \end{feynman}
\end{tikzpicture}
$$

This is the corresponding 4-point vertex we are considering, for which we will do our loop corrections , corresponding loop level diagram with all the possible loops, with different couplings.  We have used the fact that the forward and backward reaction rate are the same such that we don't have to worry about the $HH\xrightarrow{}hh$ or $hh\xrightarrow{}HH$ \\
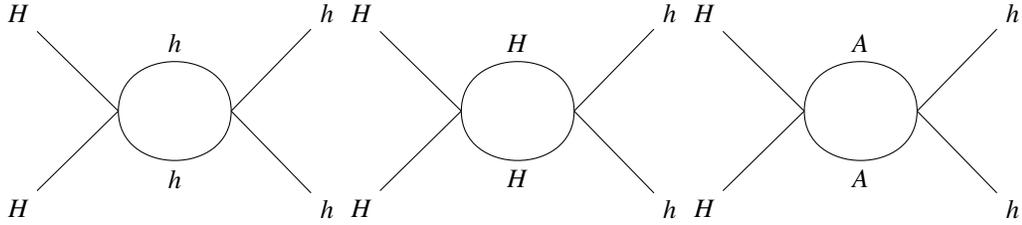
$$\begin{tikzpicture}
    \begin{feynman}
         \vertex(a);
    \vertex[ above left=of a](b){$H$};
    \vertex[ below left=of a](c){$H$};
    \vertex[right =of a](g);
    \vertex[ above right=of g](d){$h$};
    \vertex[ below right=of g](e){$h$};
    \diagram{
    (a) -- (b);
    (a) -- [half left,edge label=$h$](g);
    (g) -- [half left,edge label=$h$](a);
    (g) -- (e);
    (a) -- (c);
    (g) -- (d);
     };
    \end{feynman}
\end{tikzpicture}
\begin{tikzpicture}
    \begin{feynman}
         \vertex(a);
    \vertex[ above left=of a](b){$H$};
    \vertex[ below left=of a](c){$H$};
    \vertex[right =of a](g);
    \vertex[ above right=of g](d){$h$};
    \vertex[ below right=of g](e){$h$};
    \diagram{
    (a) -- (b);
    (a) -- [half left,edge label=$H$](g);
    (g) -- [half left,edge label=$H$](a);
    (g) -- (e);
    (a) -- (c);
    (g) -- (d);
     };
    \end{feynman}
\end{tikzpicture}\begin{tikzpicture}
    \begin{feynman}
         \vertex(a);
    \vertex[ above left=of a](b){$H$};
    \vertex[ below left=of a](c){$H$};
    \vertex[right =of a](g);
    \vertex[ above right=of g](d){$h$};
    \vertex[ below right=of g](e){$h$};
    \diagram{
    (a) -- (b);
    (a) -- [half left,edge label=$A$](g);
    (g) -- [half left,edge label=$A$](a);
    (g) -- (e);
    (a) -- (c);
    (g) -- (d);
     };
    \end{feynman}
\end{tikzpicture}$$
Now their are certain questions that can be asked firstly which diagram contributes more or is it equally among the three, answer to this question can be found out once we regularize our diagrams.
Let the initial momentum states be $|P_1,P_2\rangle$ and final states as $|P_3,P_4\rangle$ and the internal propagaters have some undefined momentum as $(k,q)$, symmetric factor for the diagram is a factor of $\frac{1}{2}$\cite{10}.
 \subsection{Higgs loop}
$$
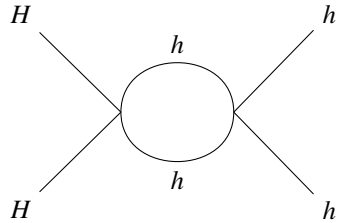
\begin{tikzpicture}
    \begin{feynman}
         \vertex(a);
    \vertex[ above left=of a](b){$H$};
    \vertex[ below left=of a](c){$H$};
    \vertex[right =of a](g);
    \vertex[ above right=of g](d){$h$};
    \vertex[ below right=of g](e){$h$};
    \diagram{
    (a) -- (b);
    (a) -- [half left,edge label=$h$](g);
    (g) -- [half left,edge label=$h$](a);
    (g) -- (e);
    (a) -- (c);
    (g) -- (d);
     };
    \end{feynman}
    \end{tikzpicture}$$
$$I_1=\frac{1}{2}(-i\lambda_{345})(-i3\lambda_1)\int \frac{i}{(k^2-m_h^2)(2\pi)^4}\int\frac{i}{(q^2-m_h^2)(2\pi)^4}$$
$$=\frac{-3\lambda_{345}\lambda_1}{128\pi^4}\int\frac{d^4k_E}{-k_E^2-m_h^2}\int\frac{d^4q_E}{-q_E^2-m_h^2}$$
$$=\frac{-3\lambda_{345}\lambda_1}{128\pi^4}\int\frac{k_E^3dk_E}{(k_E^2+m_h^2)}\int\frac{q_E^3dq_E}{q_E^2+m_h^2}$$
$$=\frac{-3\lambda_{345}\lambda_1}{128\pi^4}\int_{0}^{\infty}dy_1\frac{y_1}{y_1+m_h^2}\int_{0}^{\infty}dy_2\frac{y_2}{y_2+m_h^2}$$
$$=\frac{-3\lambda_{345}\lambda_1}{128\pi^4}\lim_{\Lambda\xrightarrow{}\infty}\int_{0}^{\Lambda^2}dy_1\frac{y_1}{y_1+m_h^2}\int_{0}^{\Lambda^2}dy_2\frac{y_2}{y_2+m_h^2}$$
$$=\frac{-3\lambda_{345}\lambda_1}{128\pi^4}\lim_{\Lambda\xrightarrow{}\infty}\left(\Lambda^2-m_h^2\ln(m_h^2+\Lambda^2)\right)\left(\Lambda^2-m_h^2\ln(m_h^2+\Lambda^2)\right)$$
$$=\frac{-3\lambda_{345}\lambda_1}{128\pi^4}\lim_{\Lambda\xrightarrow{}\infty}\left(\Lambda^2-m_h^2\ln(m_h^2+\Lambda^2)\right)^2$$
$$=\frac{-3(\lambda_3+\lambda_4+\lambda_5)(\lambda_1)}{128\pi^4}\left[(\Lambda)^4+\left(\ln(m_h^2+\Lambda^2)\right)^2-2\ln(m_h^2+\Lambda^2)\right]$$\\
Now we expand our series at infinity as Puiseux series.
$$=\frac{-3(\lambda_3+\lambda_4+\lambda_5)(\lambda_1)}{128\pi^4}\left(\Lambda^4+\left[\ln(\Lambda^2)+\frac{m_h^2}{\Lambda^2}-\frac{(m_h^2)^2}{2(\Lambda^2)^2}+\cdots O\left(\frac{1}{\Lambda^2}\right)^7\right]\right)$$
$$- \frac{-3(\lambda_3+\lambda_4+\lambda_5)(\lambda_1)}{128\pi^4}2\Lambda^2\cross\left(\left[\ln(\lambda^2)+\frac{m_h^2}{\Lambda^2}-\frac{(m_h^2)^2}{2(\Lambda^2)^2} +\cdots O\left(\frac{1}{\Lambda^2}\right)^7\right]\right)$$
\begin{align*}
    \Aboxed{I_1=\lim_{\Lambda\xrightarrow{}\infty}\frac{-3(\lambda_3+\lambda_4+\lambda_5)(\lambda_1)}{128\pi^4}\left[\Lambda^4+(\ln(\Lambda^2))^2-2\Lambda^2\ln(\Lambda^2)+2m_h^2-\frac{(m_h^2)^2}{\Lambda^2}\right]}
\end{align*}
\subsection{Other possible scalar loops}
$$
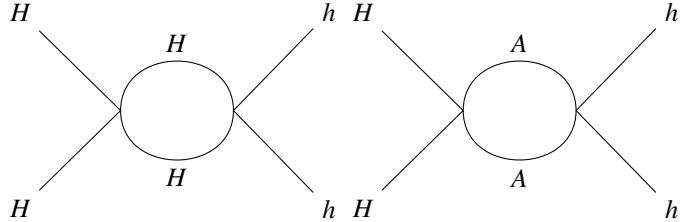
\begin{tikzpicture}
    \begin{feynman}
         \vertex(a);
    \vertex[ above left=of a](b){$H$};
    \vertex[ below left=of a](c){$H$};
    \vertex[right =of a](g);
    \vertex[ above right=of g](d){$h$};
    \vertex[ below right=of g](e){$h$};
    \diagram{
    (a) -- (b);
    (a) -- [half left,edge label=$H$](g);
    (g) -- [half left,edge label=$H$](a);
    (g) -- (e);
    (a) -- (c);
    (g) -- (d);
     };
    \end{feynman}
\end{tikzpicture}\begin{tikzpicture}
    \begin{feynman}
         \vertex(a);
    \vertex[ above left=of a](b){$H$};
    \vertex[ below left=of a](c){$H$};
    \vertex[right =of a](g);
    \vertex[ above right=of g](d){$h$};
    \vertex[ below right=of g](e){$h$};
    \diagram{
    (a) -- (b);
    (a) -- [half left,edge label=$A$](g);
    (g) -- [half left,edge label=$A$](a);
    (g) -- (e);
    (a) -- (c);
    (g) -- (d);
     }
     ;
    \end{feynman}
\end{tikzpicture}$$
$I_2$= Calculation of S-matrix for DM candidate loop\\
$I_3$= Calculation of S-matrix for other scalar loop\\
\begin{align*}
    \Aboxed{I_2=\lim_{\Lambda\xrightarrow{}\infty}\frac{-3(\lambda_3+\lambda_4+\lambda_5)(\lambda_2)}{128\pi^4}\left[\Lambda^4+(\ln(\Lambda^2))^2-2\Lambda^2\ln(\Lambda^2)+2m_H^2-\frac{(m_H^2)^2}{\Lambda^2}\right]}
\end{align*}
\begin{align*}
    \Aboxed{I_3=\lim_{\Lambda\xrightarrow{}\infty}\frac{-(\lambda_3+\lambda_4-\lambda_5)(\lambda_2)}{128\pi^4}\left[\Lambda^4+(\ln(\Lambda^2))^2-2\Lambda^2\ln(\Lambda^2)+2m_A^2-\frac{(m_A^2)^2}{\Lambda^2}\right]}   
\end{align*}\\
Its clear now that the contribution depends on the coupling values of $\lambda_1,\lambda_2,\lambda_3,\lambda_4,\lambda_5$, these values will which diagrams have a major contribution.\\
One more interesting thing to note is that that as we started out with the fact that we can choose $H$ or $A$ as the masses were almost similar for our DM candidate but now we see that they don't contribute equally to the one loop diagrams, the major contribution will come from the diagram for where we have used $HHHH$ and $hhHH$ vertex, this can be seen from the above formulas.
\subsection{Re-normalization of the theory}
We have used the method of multiplicative re-normalization for our theory, a brief discussion will be presented here about the method. We first have regularized the diagram such that we have separated out the divergence part and later we Taylor expand it around zero momentum(for massive theory). We then add the Counter terms to the Lagrangian to cancel out these divergence, the order of the C.T will be of the order of the $h$ or any other constant we have loop expanded our series\cite{8}.

The part of Lagrangian we are concerned is,
$$\mathcal{L}=-\frac{m_H^2}{2}(HH)-\frac{m_h^2}{2}(hh)+(\lambda_3+\lambda_4+\lambda_5)(HHhh)-3\lambda_2(HHHH)-3\lambda_1(hhhh)$$
$$\mathcal{L}_{fin}=-\frac{m_H^2}{2}H^2-\frac{m_h^2}{2}h^2+\lambda_{345}H^2h^2-3\lambda_2H^4-3\lambda_1h^4 + C.T$$
We haven't done the loop correction for the mass terms or the two point function we have only done the correction for 4-point vertex of $HHhh$ so when we add the counter terms only the term in Lagrangian with this vertex will be redefined as,\\
$$\mathcal{L}_{fin}=-\frac{m_H^2}{2}H^2-\frac{m_h^2}{2}h^2+\lambda_{345}H^2h^2-3\lambda_2H^4-3\lambda_1h^4-\frac{A}{2}H^2-\frac{B}{2}h^2+CH^2h^2-DH^4-Eh^4$$
$$=-\frac{(m_H^2+A)}{2}H^2-\frac{(m_h^2+B)}{2}h^2+(\lambda_{345}+C)H^2h^2-(3\lambda_2+D)H^4-(3\lambda_1+E)h^4$$
All the terms in the counter terms except for $HHhh$ vertex will be zero as, ($C_1,C_2,C_3$) are the the C.T for different loops as discussed above, 
$$A=0$$
$$B=0$$
$$C_1=\frac{-3}{128\pi^4}\left[\Lambda^4+(\ln(\Lambda^2))^2-2\Lambda^2\ln(\Lambda^2)+2m_h^2-\frac{(m_h^2)^2}{\Lambda^2}\right]$$
$$C_2=\frac{-3}{128\pi^4}\left[\Lambda^4+(\ln(\Lambda^2))^2-2\Lambda^2\ln(\Lambda^2)+2m_H^2-\frac{(m_H^2)^2}{\Lambda^2}\right]$$
$$C_3=\frac{-3}{128\pi^4}\left[\Lambda^4+(\ln(\Lambda^2))^2-2\Lambda^2\ln(\Lambda^2)+2m_A^2-\frac{(m_A^2)^2}{\Lambda^2}\right]$$
$$D=0$$
$$E=0$$
We have redefined our Lagrangian such that the theory renders finite value and as we go higher order in perturbation we have to keep redefining our parameters to encounter for divergences. 
\section*{Acknowledgement}
I would like to thank Dr Abhass Kumar\footnote{University of Delhi, Department of Physics and astrophysics} under his supervision i was able to complete this research project. And providing me with insights due to which the calculations were made possible.
%
%
%
%

\end{document}